\documentclass[11pt]{article}
\usepackage[centering]{geometry}
\usepackage{amsmath}
\usepackage{txfonts}
\usepackage{cite}
\usepackage[dvipdfmx]{hyperref}
\hypersetup{
	colorlinks=true,
	linkcolor=red,
	citecolor=blue
}
\newcommand{\one}{\boldsymbol{1}}
\newcommand{\Ncal}{\mathcal{N}}
\newcommand{\ebd}{\boldsymbol{e}}
\newcommand{\fbd}{\boldsymbol{f}}
\newcommand{\gbd}{\boldsymbol{g}}
\newcommand{\zbd}{\boldsymbol{z}}
\newcommand{\Acal}{\mathcal{A}}
\newcommand{\Rcal}{\mathcal{R}}

\makeatletter

\@addtoreset{equation}{section}
\makeatother

\begin{document}

\title{
	\vspace{-30pt}
	{\Large \bf Extended supersymmetry with central charges 
		in Dirac action \\ with curved extra dimensions
		\\*[20pt]}
}

\author{
	Inori~Ueba\footnote{E-mail address: i-ueba@stu.kobe-u.ac.jp}\\*[10pt]
	{\it \normalsize Department of Physics, Kobe University, Kobe 657-8501, Japan}\\*[55pt]
}

\date{
	\centerline{\small \bf Abstract}
	\begin{minipage}{0.9\textwidth}
		\medskip\medskip 
		\normalsize
We discuss a new realization of $\mathcal{N}$-extended quantum-mechanical supersymmetry (QM SUSY) with central charges hidden in the four-dimensional (4D) mass spectrum of higher dimensional Dirac action with curved extra dimensions. We show that this $\mathcal{N}$-extended QM SUSY results from symmetries in extra dimensions, and the supermultiplets in this supersymmetry algebra correspond to the Bogomol'nyi--Prasad--Sommerfield states. Furthermore, we examine the model of the $S^2$-extra dimension with a magnetic monopole background and confirm that the $\mathcal{N}$-extended QM SUSY explains the degeneracy 
of the 4D mass spectrum.
		\begin{flushleft}
			\hspace{10ex}  extra dimensions, curved space
		\end{flushleft}
		\begin{flushright}
			\normalsize KOBE-TH-19-05 \\*[55pt]
		\end{flushright}
	\end{minipage}}

	\begin{titlepage}
		\maketitle
		\thispagestyle{empty}
	\end{titlepage}

\section{Introduction}
\label{sec:Introduction}
So far, quantum-mechanical supersymmetry (QM SUSY) has attracted much attention and has been applied to the various research areas, e.g. exactly solvable quantum mechanics~\cite{Cooper:1994eh,Bagchi2009,Odake2009,Fernandez2009}, Berry phase~\cite{Pedder:2007ff,Ohya:2014ska,Ohya:2015xya}, black holes and AdS/CFT~\cite{claus1998,Gibbons1999,pacumio2000,Bellucci2003,Bakas2009}, Sachdev--Ye--Kitaev model~\cite{Fu2017,Li2017,Yoon2017,Murugan2017,Jones2018}, extra dimensional models~\cite{Lim:2005rc,Lim:2007fy,Lim:2008hi,Nagasawa2011,Fujimoto:2016llj,Fujimoto:2016rbr} and so on.
Its extensions are also investigated.
The $\Ncal$-extended supersymmetry is the extension which includes  $\Ncal$ independent supercharges in the supersymmetry algebra~\cite{Toppan2001,Kuznetsova2006,Faux2005,deCrombrugghe:1982nmd,Howe1988,Akulov1999,Nagasawa2004,Nagasawa2005}.
Each of supercharge corresponds to a square root of Hamiltonian, and they explain
the degeneracy of the energy spectrum.
In addition, the central extension which introduces central charges in the algebra is also studied~\cite{Ivanov:1990cz,Niederle:2001cu,Faux:2003hn,Bellucci:2005gp}.\footnote{Spontaneous generations of the central charges in field-theoretic SUSY algebras and associated materials have been discussed (see e.g.~\cite{Fayet:1975yi,Fayet:1978ig,Fayet:1984wm,Fayet:1985kt}).}
Central charges commute with all the operators in the algebra.
As is well known, if there are central charges, the size of supermultiplets can be small compared with 
the regular representation~\cite{Witten:1978mh,WessBagger}. 
Such multiplets are called short multiplets or Bogomol'nyi--Prasad--Sommerfield (BPS) states.\footnote{See also the original papers of BPS states~\cite{Prasad:1975kr,Bogomolny:1975de}.}
Since not so many models which realize arbitrary large $\Ncal$-extended QM SUSY with central charges are known, it is worth investigating a new realization of $\Ncal$-extended one.

Here, we focus on the higher dimensional Dirac action with extra dimensions. In Refs.~\cite{Fujimoto:2016llj,Fujimoto:2016rbr}, it has been shown that the structure of the $\Ncal=2$ QM SUSY is hidden in the four-dimensional (4D) mass spectrum of the higher dimensional Dirac action with flat extra dimensions, and the Kaluza-Klein (KK) mode functions for the 4D right-handed and left-handed spinors form the supermultiplets. 
Furthermore, in the previous papers~\cite{Fujimoto:2018cnf,Fujimoto:2018tjm}, we have revealed that this $\Ncal=2$ QM SUSY can be extended to the $\Ncal$-extended QM SUSY from the reflection symmetries in the flat extra dimensions. Then we have found that the central charges appear as the result of the reflection symmetries, and the supermultiplets of this extended QM SUSY corresponds to the BPS states.
These supercharges can explain the degeneracy of the 4D mass spectrum.

However, there remain many tasks to be addressed. Here we focus on the following ones: First, previous works are only devoted to the case of the flat extra dimensions. Therefore we should take into account the case of curved extra dimensions for a general discussion. Second, we have only considered the $\Ncal$-extended QM SUSY from the reflection symmetries. If there are more symmetries in extra dimensions,  additional degeneracies would appear in 4D mass spectra. Then, we can expect that further structures of $\Ncal$-extended QM SUSY from other symmetries are hidden in them.

Based on the above, in this paper, we discuss a new realization of the $\Ncal$-extended QM SUSY with central charges which are obtained from symmetries in the higher dimensional Dirac action with curved extra dimensions.
We show that the central charges appear from those symmetries and this $\Ncal$-extended QM SUSY corresponds to the generalization of the previous one.
Furthermore, the supermultiplets in this SUSY algebra also become the BPS states.
Then, as an example, we will confirm that this $\Ncal$-extended QM SUSY 
is realized in the $S^2$-extra dimension with the magnetic monopole background
and explain the degeneracy of the 4D mass spectrum.

This paper is organized as follows: In Section \ref{sec:N=2-SUSY}, we summarize the KK decomposition of the $(4+d)$-dimensional Dirac field with the curved extra dimension and show that the $\Ncal=2$
QM SUSY is hidden in the 4D mass spectrum. 
In Section \ref{sec:N-extended}, we construct the $\Ncal$-extended QM SUSY with central charges from the symmetries in extra dimensions and discuss the representation of this SUSY algebra.
Then, in Section \ref{sec:exapmle}, we confirm that this $\Ncal$-extended QM SUSY can be realized in the 
model of the $S^2$-extra dimension with the magnetic monopole background, and the KK mode functions 
correspond to the representation given in Section \ref{sec:N-extended}. 
Section \ref{sec:summary} is devoted to summary and discussion.

%
\section{$\Ncal=2$ QM SUSY in higher dimensional Dirac action}
\label{sec:N=2-SUSY}
%
In this section, we show that the structure of $\Ncal=2$ QM SUSY is always hidden in the 4D mass spectrum of the $(4+d)$-dimensional Dirac action with curved extra dimensions.

First, we assume that the $(4+d)$-dimensional metric $G_{MN}\ (M,N=0,1,2,3,y_1,\cdots ,y_d)$ is of the form
	\begin{align}
		ds^2=G_{MN}dx^M dx^N
		=e^{2\Delta(y)}\eta_{\mu\nu}dx^\mu dx^\nu+g_{yy'}(y)\,dy^{y}dy^{y'}\,,
	\end{align}
where $x^M=(x^\mu,y^{y})$ is the $(4+d)$-dimensional coordinates, and then $x^\mu\ (\mu=0,1,2,3)$ and $y^{y}\ (y=y_1\cdots y_d)$ indicate the coordinates of the 4D and the extra dimensional space, respectively.
$\eta_{\mu\nu}=\rm{diag}(-1,+1,+1,+1)$ denotes the 4D Minkowski metric, and $\Delta(y)$ and $g_{yy'}(y)$ depend only on the extra dimensional coordinates.\footnote{In the case that $\Delta(y)=-k|y|$ and $g_{yy}=1$ with the 1d extra dimension, this metric corresponds to the Randall-Sundrum warped metric~\cite{Randall:1999ee,Randall:1999vf}.}

Then, for a general discussion, we study the ($4+d$)-dimensional Dirac action with the vector background field $A_N(y)=(0,A_{y}(y))$ and the  scalar background field $W(y)$:
	\begin{align}
		S=\int d^4x\int_\Omega d^dy \sqrt{-G}\, \bar\Psi(x,y)\left[i\Gamma^{\hat{N}}{e_{\hat{N}}}^N\left(\nabla_N+iqA_N(y)\right)-W(y)\right]\Psi(x,y)\,,
	\end{align}
where $\Omega$ represents the space of the extra dimensions, $G=\det{G_{MN}}$ and we define the Dirac conjugate as $\bar\Psi(x,y)=\Psi^\dagger(x,y)\Gamma^{\hat{0}}\,$.
$\Gamma^{\hat{N}}$ indicate the gamma matrices which are defined by
	\begin{align}
		\{\Gamma^{\hat{M}}\,,\Gamma^{\hat{N}}\}=-2\eta^{\hat{M}\hat{N}}\one_{2^{\lfloor d/2\rfloor+2}}\,,
		&&
		(\Gamma^{\hat{M}})^\dagger=\Gamma^{\hat{0}}\Gamma^{\hat{M}}\Gamma^{\hat{0}}
		&&
		(\hat{M},\hat{N}\,=\hat{0},\hat{1},\hat{2},\hat{3},\hat{y}_1,\cdots,\hat{y}_d)\,,
	\end{align} 
where $\hat{M},\hat{N}$ denote the indices of the local Lorentz frame, and $\eta_{\hat{M}\hat{N}}=\rm{diag}(-1,+1,\cdots,+1)\,$ is $(4+d)$-dimensional Minkowski metric.
${e_N}^{\hat N}$ is the vielvein and satisfies
\begin{align}
{e_L}^{\hat M}{e_{\hat M}}^N=\delta_{L}^N\,,&&{e_{\hat L}}^M{e_M}^{\hat N}=\delta_{\hat L}^{\hat N}\,,&&G_{MN}={e_M}^{\hat M}{e_N}^{\hat N}\eta_{\hat{M}\hat{N}}\,.
\end{align}
In this model, the nonzero components of the vielvein can be given by ${e_\mu}^{\hat\mu}=e^{\Delta(y)}\delta_\mu^{\hat\mu}\  (\mu=0,1,2,3\,,\ \hat\mu=\hat{0},\hat{1},\hat{2},\hat{3})$ and ${e_{y}}^{\hat{y}}\ (y=y_1\cdots y_d\,,\ \hat{y}=\hat{y}_1,\cdots,\hat{y}_d)$. 
$\nabla_N$ represents the covariant derivative whose behavior for the Dirac field $\Psi(x,y)$ is
	\begin{align}
		\nabla_N\Psi(x,y)=\left(\partial_N+\frac{i}{2}\omega_{N\hat{K}\hat{L}}\Sigma^{\hat{K}\hat{L}}\right)\Psi(x,y)\,,
	\end{align}
where $\Sigma^{\hat{K}\hat{L}}=\frac{i}{4}[\Gamma^{\hat{K}}\,,\Gamma^{\hat{L}}]$ corresponds to the generator of $(4+d)$-dimensional Lorentz transformation, and $\omega_{N\hat{K}\hat{L}}$ is the spin connection defined from the Christoffel symbol $\Gamma^L_{NK}$
	\begin{align}
		{{\omega_N}^{\hat{K}}}_{\hat{L}}&=-{e_{\hat{L}}}^K\partial_N{e_K}^{\hat{K}}+{e_{\hat{L}}}^K\Gamma^L_{NK}{e_L}^{\hat{K}}\,,
	\end{align}
and whose nonzero components are $\omega_{\mu\hat\nu\hat{y}}
=\frac{1}{2}\left(e^{-\Delta}\partial_{y}e^{2\Delta} \right)\eta_{\mu\hat\nu}{e_{\hat{y}}}^{y}$ and $\omega_{y\hat{y}'\hat{y}''}$ in this model.
	
For the convenience, we adopt the following representation of the gamma matrices:
	\begin{align}
		\Gamma^{\hat\mu}=\one_{2^{\lfloor d/2\rfloor}}\otimes\gamma^{\hat\mu}\,,&&
		\Gamma^{\hat{y}}=\gamma^{\hat{y}}\otimes\gamma^5\,.
		\label{eq:gamma_rep}
	\end{align}
$\gamma^{\hat\mu}\ (\hat{\mu}=\hat{0},\hat{1},\hat{2},\hat{3})$ are the $4\times4$ 4D gamma matrices and $\gamma^5=i\gamma^{\hat{0}}\gamma^{\hat{1}}\gamma^{\hat{2}}\gamma^{\hat{3}}$ denotes the 4D chiral matrix. $\gamma^{\hat{y}}\ (\hat{y}=\hat{y}_1,\cdots,\hat{y}_d)$ represent the $2^{\lfloor d/2\rfloor}\times2^{\lfloor d/2\rfloor}$ $d$-dimensional internal gamma matrices which satisfy
	\begin{align}
		\{\gamma^{\hat{y}}\,,\gamma^{\hat{y}'}\}=-2\delta^{\hat{y}\hat{y}'}\,,\ \ \ \ (\gamma^{\hat{y}})^\dagger=-\gamma^{\hat{y}}\,.
	\end{align}
From this representation, the Dirac operator can be rewritten into the form\footnote{$\gamma^\mu$ implies $\gamma^\mu=\gamma^{\hat\mu}\delta_{\hat\mu}^\mu$, where $\delta_{\hat\mu}^\mu$ appear from the vielvein.}
	\begin{align}
		&i\Gamma^{\hat{N}}{e_{\hat{N}}}^N\left(\nabla_N+iqA_N\right)-W
		=\one_{2^{\lfloor d/2\rfloor}}\otimes e^{-\Delta}i\gamma^{\mu}\partial_{\mu}+i\gamma^{\hat{y}}{e_{\hat{y}}}^y(\nabla_y+iqA_y+2\partial_y\Delta)\otimes\gamma^5-W\one_{2^{\lfloor d/2\rfloor}}\otimes\one_4\,,
		\label{eq:Dirac_opertor}
	\end{align}
where $\nabla_y$ in the right hand side of \eqref{eq:Dirac_opertor} means
	\begin{align}
		\nabla_y=\partial_y+\frac{i}{2}\omega_{y\hat{y}'\hat{y}''}\sigma^{\hat{y}'\hat{y}''}\,,&& \sigma^{\hat{y'}\hat{y''}}=\frac{i}{4}\left[\gamma^{\hat y'}\,,\gamma^{\hat y''}\right]\,,
	\end{align}
and this corresponds to the covariant derivative for spinors defined on the space $\Omega$.

Next, we consider the KK decomposition of the higher dimensional Dirac field $\Psi(x,y)$ to obtain the action with 4D fields:
	\begin{align}
		\Psi(x,y)
		& =\sum_{n}\sum_{\alpha}
		\Big\{\,e^{-2\Delta(y)}\boldsymbol{f}_{\alpha}^{(n)}(y)\otimes\psi_{R,\alpha}^{(n)}(x)
		+e^{-2\Delta(y)}\boldsymbol{g}_{\alpha}^{(n)}(y) 
		\otimes\psi_{L,\alpha}^{(n)}(x) \,\Big\}\,,
		\label{eq:KK_decomposition}
	\end{align}
where the index $n$ denotes the $n$-th level of the KK modes and $\alpha$ indicates the additional degeneracy of the $n$-th KK modes (if exists).
The mode functions $\boldsymbol{f}_{\alpha}^{(n)}(y)$ ($\boldsymbol{g}_{\alpha}^{(n)}(y)$)
have $2^{\lfloor d/2\rfloor}$ components and are assumed to form a complete set
with respect to the internal space associated with the 4D right-handed
(left-handed) chiral spinors $\psi_{R,\alpha}^{(n)}(x)$ ($\psi_{L,\alpha}^{(n)}(x)$). 
By substituting \eqref{eq:KK_decomposition} and \eqref{eq:Dirac_opertor} into the action, 
we obtain
	\begin{align}
		S=\sum_{n,m}\sum_{\alpha,\beta}\int d^4x\, \bigg[
		&\langle\fbd^{(n)}_\alpha|\fbd^{(m)}_\beta\rangle\bar\psi^{(n)}_{R,\,\alpha}(x)\,i\gamma^{\mu}\partial_\mu\psi^{(m)}_{R,\,\beta}(x)
		+\langle\gbd^{(n)}_\alpha|\gbd^{(m)}_\beta\rangle\bar\psi^{(n)}_{L,\,\alpha}(x)\,i\gamma^\mu\partial_\mu\psi^{(m)}_{L,\,\beta}(x)
		\notag\\
		&-\langle\fbd^{(n)}_\alpha|\Acal^\dagger\gbd^{(m)}_\beta\rangle\bar\psi^{(n)}_{R,\,\alpha}(x)\,\psi^{(m)}_{L,\,\beta}(x)
		-\langle\gbd^{(n)}_\alpha|\Acal\fbd^{(m)}_\beta\rangle\bar\psi^{(n)}_{L,\,\alpha}(x)\,\psi^{(m)}_{R,\,\beta}(x)
		\bigg]\,,
		\label{eq:decomposition_action}
	\end{align}
where we have defined the inner product and the operator $\Acal,\Acal^\dagger$ as
	\begin{align}
		\langle X| Y\rangle&=\int_{\Omega} d^dy\sqrt{g}e^{-\Delta(y)} X^\dagger(y)\,  Y(y)\,,\ \ \ \ \ g=\det{g_{yy'}}\,,
		\\
		\mathcal{A}&=e^{\Delta}\left[-i\gamma^{\hat{y}}{e_{\hat{y}}}^y(\nabla_y+iqA_y)+W\right]\,,
		\\
		\Acal^\dagger&=e^{\Delta}\left[+i\gamma^{\hat{y}}{e_{\hat{y}}}^y(\nabla_y+iqA_y)+W\right]\,.
		\label{eq:def_A}
	\end{align}
Then, by requiring that the KK mode functions satisfy the orthonormal relations
	\begin{align}
	&\langle\fbd^{(n)}_\alpha|\fbd^{(m)}_\beta\rangle=\langle\gbd^{(n)}_\alpha|\gbd^{(m)}_\beta\rangle=\delta^{nm}\delta_{\alpha\beta}\,,
		\notag\\
		&\langle\fbd^{(n)}_\alpha|\Acal^\dagger\gbd^{(m)}_\beta\rangle=\langle\gbd^{(n)}_\alpha|\Acal\fbd^{(m)}_\beta\rangle=m_n\delta^{nm}\delta_{\alpha\beta}\,,
		\label{eq:orthonormal relations}
	\end{align}
we can obtain the following action:
	\begin{align}
		S=\int d^4x\,\bigg\{\,&\sum_{\alpha}\sum_{n}\bar{\psi}_{\alpha}^{(n)}(x)(i\gamma^{\mu}\partial_{\mu}-m_{n})\psi_{\alpha}^{(n)}(x)\nonumber\\
		&+\sum_{\alpha}\bar{\psi}_{L,\alpha}^{(0)}(x)i\gamma^{\mu}\partial_\mu\psi_{L,\alpha}^{(0)}(x)
		+\sum_{\alpha}\bar{\psi}_{R,\alpha}^{(0)}(x)i\gamma^{\mu}\partial_\mu\psi_{R,\alpha}^{(0)}(x)\,\bigg\}\,,
		\label{eq:4D_action}
	\end{align}
where $\psi_{\alpha}^{(n)}(x)=\psi_{R,\alpha}^{(n)}(x)+\psi_{L,\alpha}^{(n)}(x)$ indicate 4D Dirac spinors with mass $m_n$ and $\psi_{L/R,\alpha}^{(0)}(x)$ are massless 4D chiral spinors.
The expression of the above effective 4D action coincides with the case of flat extra dimensions given in~\cite{Fujimoto:2018tjm}, although the effects of curved spaces and background fields appear as the mass spectrum through the definition of $\Acal,\Acal^\dagger$ and \eqref{eq:orthonormal relations}.

Since we have assumed that the KK mode functions $\fbd^{(n)}_\alpha$ and $\gbd^{(n)}_\alpha$ form the complete set respectively,
the orthonormal relations \eqref{eq:orthonormal relations} lead to 
	\begin{align}
		\Acal\fbd^{(n)}_\alpha(y)=m_n\gbd^{(n)}_\alpha(y)\,,\ \ \ \Acal^\dagger\gbd^{(n)}_\alpha(y)=m_n\fbd^{(n)}_\alpha(y)\,.
	\end{align}
From the above relations, we can obtain 
	\begin{align}
		Q\begin{pmatrix}
		\boldsymbol{f}_{\alpha}^{(n)}(y) \\ 0
		\end{pmatrix}=m_n
		\begin{pmatrix}
		0 \\ \boldsymbol{g}_{\alpha}^{(n)}(y)
		\end{pmatrix}\,,&&
		H\begin{pmatrix}
		\boldsymbol{f}_{\alpha}^{(n)}(y) \\ 0
		\end{pmatrix}=m_n^2\begin{pmatrix}
		\boldsymbol{f}_{\alpha}^{(n)}(y) \\ 0
		\end{pmatrix}\,,&&
		(-1)^F\begin{pmatrix}
		\boldsymbol{f}_{\alpha}^{(n)}(y) \\ 0
		\end{pmatrix}=+\begin{pmatrix}
		\boldsymbol{f}_{\alpha}^{(n)}(y) \\ 0
		\end{pmatrix}\,,
		\notag\\
		Q\begin{pmatrix}
		0 \\ \boldsymbol{g}_{\alpha}^{(n)}(y)
		\end{pmatrix}=m_n
		\begin{pmatrix}
		\boldsymbol{f}_{\alpha}^{(n)}(y) \\ 0
		\end{pmatrix}\,,&&
		H\begin{pmatrix}
		0 \\ \boldsymbol{g}_{\alpha}^{(n)}(y)
		\end{pmatrix}=m_n^2\begin{pmatrix}
		0 \\ \boldsymbol{g}_{\alpha}^{(n)}(y)
		\end{pmatrix}\,,&&
		(-1)^F\begin{pmatrix}
		0 \\ \boldsymbol{g}_{\alpha}^{(n)}(y)
		\end{pmatrix}=-\begin{pmatrix}
		0 \\ \boldsymbol{g}_{\alpha}^{(n)}(y)
		\end{pmatrix}\,,
		\label{eq:SUSY_relation}
	\end{align}
where the supercharge $Q$, the Hamiltonian $H$ and the ``fermion'' number operator $(-1)^F$ are defined by
	\begin{align}
		Q&=\begin{pmatrix} 0  & \Acal^\dagger \\ \Acal & 0 \end{pmatrix}\,,
		\label{eq:supercharge}
		\\
		H&=Q^2
		=e^{2\Delta}\left[
		-(\nabla_y+iqA_y)^2+q\sigma^{\hat{y}\hat{y}'}{e_{\hat{y}}}^y{e_{\hat{y}'}}^{y'}F_{yy'}+\frac{1}{4}R+i\gamma^{\hat{y}}{e_{\hat{y}}}^y(\partial_yW)(-1)^F+W^2
		\right]\notag
		\\
		&\hspace{2cm}+ie^{2\Delta}\gamma^{\hat{y}}{e_{\hat{y}}}^y(\partial_y\Delta)
		\begin{pmatrix}
		\Acal & 0 \\ 0 & -\Acal^\dagger
		\end{pmatrix}\,,							
		\label{eq:Hamiltonian}
		\\
		(-1)^F&=\begin{pmatrix}
			\one_{2^{\lfloor d/2 \rfloor}} & 0 \\ 0 & -\one_{2^{\lfloor d/2 \rfloor}}
		\end{pmatrix}\,.
	\end{align}
$F_{yy'}$  is the field strength for $A_y$ and $R$ is the Ricci scalar defined on $\Omega$.
Then, we can find that the relations \eqref{eq:SUSY_relation} realize the $\Ncal=2$ supersymmetric quantum mechanics~\cite{Witten1981,Cooper:1994eh}.\footnote{The $\Ncal=2$ SUSY algebra $\{Q_{i},Q_{j}\}=2H\delta_{ij}$ $(i,j=1,2)$ is obtained with $Q_{1}=Q$ and $Q_{2}=i(-1)^{F}Q$.}
In this model, the ``bosonic'' 
and ``fermionic'' states which form an $\Ncal=2$ supermultiplet correspond to the KK mode functions
$(\boldsymbol{f}_{\alpha}^{(n)}(y),0)^{\text{T}}$ and $(0,\boldsymbol{g}_{\alpha}^{(n)}(y))^{\text{T}}$.

Before closing this section, we comment about the Hermiticity of the supercharge.
From the action principle $\delta S=0$, we obtain the following condition for the KK mode functions:
	\begin{align}
		\int_{\partial\Omega}& d^{n-1}y\sqrt{g}(\fbd_\alpha^{(n)}(y))^\dagger \,in_y(y)\gamma^{\hat{y}}{e_{\hat{y}}}^y\gbd_\beta^{(m)}(y)=0\,,
		\label{eq:generic BC}
	\end{align}
for all $m,n,\alpha,\beta$, where
$\partial\Omega$ represents the boundary of $\Omega$, and $n_{y}(y)$ is an orthonormal vector on $\partial\Omega$.
We can show that the above condition corresponds to the Hermiticity condition for the supercharge.
Then, the supercharge $Q$ is Hermitian as long as the action principle is required.
Thus, we can conclude that the $\Ncal=2$ QM SUSY is always realized in the 4D mass spectrum of the higher dimensional Dirac action and the doubly degenerate states $(\boldsymbol{f}_{\alpha}^{(n)}(y),0)^{\text{T}}$ and $(0,\boldsymbol{g}_{\alpha}^{(n)}(y))^{\text{T}}$ are mutually related 
by the supercharge $Q$, except for zero energy states.

\section{$\Ncal$-extended QM SUSY with central charges}
\label{sec:N-extended}
In the previous section, we have described the $\Ncal=2$ QM SUSY hidden in the doubly degeneracy of $\boldsymbol{f}_{\alpha}^{(n)}$ and $\boldsymbol{g}_{\alpha}^{(n)}(y)$.
However, we can expect that further hidden structures exist in the 4D mass spectrum and 
this would lead to the extra degeneracy due to the index $\alpha$ in addition to the doubly one.

In this section, we show that
the $\Ncal$-extended QM SUSY with central charges can be constructed from symmetries in the extra dimensions. This QM SUSY can explain the extra degeneracy in the 4D mass spectrum.
Then, we clarify the representation of this algebra for the nonzero energy states and it will turn out that the eigenstates become BPS states.
This section is devoted to the general discussion, and a concrete example will be given in the next section.

\subsection{$\Ncal$-extended SUSY algebra with central charges}

Here, we discuss a new realization of $\Ncal$-extended QM SUSY from symmetries.
First, we consider sets of operators $\{\hat{a}_i\ (i=1,2,\cdots,N_a)\},\ \{\hat{b}_i\ (i=1,2,\cdots,N_b)\},\ \cdots,\ \{\hat{\alpha}_i\ (i=1,2,\cdots,N_\alpha)\},\ \{\hat{\beta}_i\ (i=1,2,\cdots,N_\beta)\},\ \cdots$, which are Hermitian and consistent with an imposed boundary condition for the mode functions $\fbd_\alpha^{(n)}$,\footnote{More precisely, we require that the functions $\hat{a}_i\fbd_\alpha^{(n)}\ (i=1,2,\cdots,N_a)$, $\hat{b}_i\fbd_\alpha^{(n)}\ (i=1,2,\cdots,N_b)$, $\cdots$ also satisfy the imposed boundary condition.} and commute with $\Acal^\dagger\Acal$
	\begin{align}
		&[\hat{a}_i\,,\Acal^\dagger\Acal]=[\hat{b}_i\,,\Acal^\dagger\Acal]=\cdots=[\hat{\alpha}_i\,,\Acal^\dagger\Acal]=[\hat{\beta}_i\,,\Acal^\dagger\Acal]=\cdots=0\,.
		\label{eq:commutation_relation_for_sets_of_operators}
	\end{align}
Therefore, these operators do not change the mass eigenvalues and would be related to the symmetries in the extra dimensions.
Furthermore, we require that these operators commute with the ones in the same sets
	\begin{align}
		&[\hat{a}_i\,,\hat{a}_j]=[\hat{b}_i\,,\hat{b}_j]=\cdots=0\,,&&[\hat{\alpha}_i\,,\hat{\alpha}_j]=[\hat{\beta}_i\,,\hat{\beta}_j]=\cdots=0\,,
	\end{align}
and anticommute with the ones in the different sets for Roman and Greek letters
	\begin{align}
		&\{\hat{a}_i\,,\hat{b}_j\}=\{\hat{a}_i\,,\hat{c}_j\}=\cdots=\{\hat{b}_i\,,\hat{c}_j\}=\{\hat{b}_i\,,\hat{d}_j\}=\cdots=0\,,
		\\ &\{\hat{\alpha}_i\,,\hat{\beta}_j\}=\{\hat{\alpha}_i\,,\hat{\gamma}_j\}=\cdots=\{\hat{\beta}_i\,,\hat{\gamma}_j\}=\{\hat{\beta}_i\,,\hat{\delta}_j\}=\cdots=0\,,
	\end{align}
and the operators with the Roman letters and the ones with the Greek letters commute with each other
	\begin{align}
		[\hat{a}_i\,,\hat{\alpha}_j]=[\hat{a}_i\,,\hat{\beta}_j]=\cdots=[\hat{b}_i\,,\hat{\alpha}_j]=[\hat{b}_i\,,\hat{\beta}_j]=\cdots=0\,.
	\end{align}
	
Then, we define the following extended supercharges
	\begin{align}
		&Q^{(a)}_i=\begin{pmatrix}
			& i
			\hat{a}_i\Acal^\dagger \\ -i\Acal\hat{a}_i & 
		\end{pmatrix}\,,&
		&Q^{(b)}_i=\begin{pmatrix}
			& i
			\hat{b}_i\Acal^\dagger \\ -i\Acal\hat{b}_i & 
		\end{pmatrix}\,,&
		&\cdots\,,&
		\\
		&Q^{(\alpha)}_i=\begin{pmatrix}
			& 
			\hat{\alpha}_i\Acal^\dagger \\ \Acal\hat{\alpha}_i & 
		\end{pmatrix}\,,&
		&Q^{(\beta)}_i=\begin{pmatrix}
			& 
			\hat{\beta}_i\Acal^\dagger \\ \Acal\hat{\beta}_i & 
		\end{pmatrix}\,,&
		&\cdots\,,&
	\end{align}
and obtain $\Ncal=(N_a+N_b+\cdots+N_\alpha+N_\beta+\cdots)$ SUSY algebra with the central charges\footnote{Although we can also define the supercharges with the replacement of the operators with the Roman and the Greek letters, those are essentially same as the ones given in the above.}
	\begin{align}
		&\{Q^{(A)}_i\,,Q^{(B)}_j\}=2H\delta_{ij}\delta^{AB}+2Z^{(A)}_{ij}\delta^{AB}\,,
		\label{eq:gen_SUSY_alg}
		\\
		&[Q^{(A)}_i\,,Z^{(B)}_{jk}]=[H\,,Z^{(A)}_{jk}]=[Z^{(A)}_{ij}\,,Z^{(B)}_{kl}]=[Q^{(A)}_i\,,H]=0
		\ \ \ \ (A,B=a,b,\cdots,\alpha,\beta,\cdots)\,,
	\end{align}
where $H$ denotes the Hamiltonian given by \eqref{eq:Hamiltonian} and the central charges $Z^{(A)}_{ij}$ are defined as
	\begin{align}
		Z^{(a)}_{ij}=-H\delta_{ij}+\begin{pmatrix}
			\hat{a}_i\hat{a}_j\Acal^\dagger\Acal & 0 \\ 0 & \Acal\hat{a}_i\hat{a}_j\Acal^\dagger
		\end{pmatrix}\,,
		&&
		Z^{(b)}_{ij}=-H\delta_{ij}+\begin{pmatrix}
			\hat{b}_i\hat{b}_j\Acal^\dagger\Acal & 0 \\ 0 & \Acal\hat{b}_i\hat{b}_j\Acal^\dagger
		\end{pmatrix}\,,
		&&
		\cdots\,,
		\\
		Z^{(\alpha)}_{ij}=-H\delta_{ij}+\begin{pmatrix}
			\hat{\alpha}_i\hat{\alpha}_j\Acal^\dagger\Acal & 0 \\ 0 & \Acal\hat{\alpha}_i\hat{\alpha}_j\Acal^\dagger
		\end{pmatrix}\,,
		&&
		Z^{(\beta)}_{ij}=-H\delta_{ij}+\begin{pmatrix}
			\hat{\beta}_i\hat{\beta}_j\Acal^\dagger\Acal & 0 \\ 0 & \Acal\hat{\beta}_i\hat{\beta}_j\Acal^\dagger
		\end{pmatrix}\,,
		&&
		\cdots\,.
	\end{align}
Therefore, we can consider that the central charges in this SUSY algebra result from the symmetries in the extra dimensions.
If we take the sets of operators as reflection operators and gamma matrices, this extended QM SUSY corresponds the one given in the previous papers~\cite{Fujimoto:2018tjm}.

For the existence of the extended SUSY with given sets of operators, the metric of curved spaces and the background fields are restricted to satisfy the condition \eqref{eq:commutation_relation_for_sets_of_operators}.
However, it seems difficult to find the constraints without any assumption for sets of operators.\footnote{In the Refs.~\cite{Kirchberg:2004za,Ivanov:2010ki}, 
the structure of the extended QM SUSY {\it without} central charges is discussed,
which consists of the Dirac operator in diverse dimensions on curved spaces with background gauge fields.
They have introduced tensor fields to extend the Dirac operator and constructed the supercharges, and shown the strict constraints on the geometry and the gauge fields. However, since the way of our extension is different from them and our QM SUSY admits central charges, the constraints on the geometry and the background fields will be also different from their case.
} 
Therefore, we will first prepare the geometry and the background fields, and then consider the sets of operators consistent with them when we see an example in section \ref{sec:exapmle}.

It should be mentioned that this central extension is given by direct sums of mutually (anti)commuting $\Ncal=2$ SUSY algebras as well as the previous paper~\cite{Fujimoto:2018tjm}, with different ``Hamiltonians'' for each of them.
Especially, the supercharges $Q^{(A)}_i\ (i=1,\cdots,N_A)$ for each index $A$ commute with each other 
	\begin{align}
		[Q^{(A)}_i\,,Q^{(A)}_j]=0\ \ \ \ \ (i,j=1,\cdots,N_A)\,.
	\end{align}
This property is important for the discussion of the BPS states in the next subsection.

\subsection{Representation of SUSY algebra}
\label{subsec:rep_ex_SUSY}
Then, let us clarify the representation of this algebra for the nonzero energy states.
Since the Hamiltonian and the central charges commute with each other, we first look at the simultaneous eigenstates of them:
	\begin{align}
		H\Phi^{(n)}_{\boldsymbol{s},\zbd}=m_n^2\Phi^{(n)}_{\boldsymbol{s},\zbd}\,,
		&&Z_{ij}^{(A)}\Phi^{(n)}_{\boldsymbol{s},\zbd}=z_{ij}^{(A)}m_n^2\Phi^{(n)}_{\boldsymbol{s},\zbd}\,,
	\end{align}
where $n$ and $\zbd$ indicate the labels of their eigenvalues $m_n$ and $z_{ij}^{(A)}$,\footnote{
In general, $z_{ij}^{(A)}$ depend on the index $n$ although we do not explicitly label $z_{ij}^{(A)}$ with it.} and $\boldsymbol{s}$ denotes the extra index to further classify the eigenstates in the following discussions.
For these states, 
the algebra \eqref{eq:gen_SUSY_alg} is rewritten into
	\begin{align}
		\{Q^{(A)}_i\,,Q^{(B)}_j\}=2m_n^2\delta_{ij}\delta^{AB}+2z^{(A)}_{ij}m_n^2\delta^{AB}\,.
	\end{align}
Since $z^{(A)}_{ij}$ is the real symmetric matrix, it can be diagonalized by the orthogonal matrix $U^{(A)}_{ij}$. Then, by redefining the supercharges,
we can obtain
	\begin{align}
		\{Q'^{(A)}_i\,,Q'^{(B)}_j\}=2m_n^2\delta_{ij}\delta^{AB}+2z'^{(A)}_{i}m_n^2\delta_{ij}\delta^{AB}\,,
		\label{eq:redfined_alg}
	\end{align}
where $z'^{(A)}_{i}\delta_{ij}=\sum_{k,l}U^{(A)}_{ik}z^{(A)}_{kl}({U^{(A)}})^T_{lj}$ is the diagonalized matrix and $Q'^{(A)}_i=\sum_{k}U^{(A)}_{ik}Q^{(A)}_k\,,\ Q'^{(B)}_j=\sum_{k}U^{(B)}_{jk}Q^{(B)}_k$ indicate the redefined supercharges.
Then we can find that the square of the supercharge $(Q'^{(A)}_i)^2$ equals to 0 for eigenstates with $z'^{(A)}_{i}=-1$ (if exist).
This relation leads to $Q'^{(A)}_i=0$ for such states since the supercharge $Q'^{(A)}_i$ is Hermitian.
Thus, for states with $z'^{(A)}_{i}=-1$, the number of nontrivial supercharges effectively reduces and the size of supermultiplets become small compared with the regular representation.
These states are called BPS states. 

It should be noted that, in this algebra, at most one supercharge among $Q'^{(A)}_i\ (i=1,\cdots,N_A)$ can become nontrivial and the others must equal to 0 for any eigenstates. This is because
the supercharges satisfy the relation
\begin{align}
Q'^{(A)}_iQ'^{(A)}_j=0\ \ \ (i\neq j)
\end{align}
due to \eqref{eq:redfined_alg} and the commutativity of $Q^{(A)}_i$ and $Q^{(A)}_j$.
Therefore, the number of nontrivial supercharges for the eigenstates are maximally given by the number of the sets of the operators related to symmetries, and the eigenstates necessarily become the BPS states if the sets have more than one operator.

Next, to construct the supermultiplets in this SUSY algebra, we introduce the following operators by means of the nontrivial redefined supercharges for the eigenstates:
	\begin{align}
		S_{ij}^{(AB)}=-iQ'^{(A)}_{i}Q'^{(B)}_{j}\,,
		\ \ \ \ \ 
		S_{kl}^{(CD)}=-iQ'^{(C)}_{k}Q'^{(D)}_{l}\,,
		\ \ \ \ \
		 \cdots\,,
	\end{align}
where $A,B,C,$ and $D$ are different with each other according to the above discussion. These operators commute with each other, and
therefore, we can further classify the eigenfunctions by these operators.
Since these operators satisfy
	\begin{align}
		(S_{ij}^{(AB)})^2=m_n^4(1+z_{i}'^{(A)})(1+z_{j}'^{(B)})
	\end{align}
for the eigenstates,
we can parametrize their eigenvalues as\footnote{$z_{i}'\geq-1$ because $(Q^{(A)}_i)^2=m_n^2+z'^{(A)}_{i}m_n^2\geq0$.}
	\begin{align}
		S_{ij}^{(AB)}\Phi^{(n)}_{s_{ij}^{(AB)}s_{kl}^{(CD)}\cdots,\zbd}=s_{ij}^{(AB)}\sqrt{(1+z_{i}'^{(A)})(1+z_{j}'^{(B)})}\,m_n^2\Phi^{(n)}_{s_{ij}^{(AB)}s_{kl}^{(CD)}\cdots,\zbd}\,,
	\end{align}
where $s_{ij}^{(AB)}=\pm$.
Here, we have described the index $\boldsymbol{s}$ as $s_{ij}^{(AB)}s_{kl}^{(CD)}\cdots$.

From the relation
	\begin{align}
		S_{ij}^{(AB)}Q_{m}'^{(E)}=
		\begin{cases}
			-Q_m'^{(E)}S_{ij}^{(AB)} & (E=A\,,\ m=i\ \text{or}\ E=B\,,\ m=j)\,,
			\\
			+Q_m'^{(E)}S_{ij}^{(AB)} & (\text{the others})\,,
		\end{cases}
	\end{align}
we can see that $Q_{i}'^{(A)}$ and $Q_{j}'^{(B)}$ flip the sign of the eigenvalue of $S_{ij}^{(AB)}$ but do not change other eigenvalues.
Thus, if there are $K$ nontrivial supercharges,
this relation implies that a supermultiplet composes of the $2^{\lfloor K/2 \rfloor}$-fold degenerate states $\big\{\Phi^{(n)}_{s_{ij}^{(AB)}s_{kl}^{(CD)}\cdots,\zbd}\ \text{with}\  s_{ij}^{(AB)}=\pm\,,\ s_{kl}^{(CD)}=\pm\,,\ \cdots\big\}$ and we can explicitly construct the supermultiplet from $\Phi^{(n)}_{++\cdots,\zbd}$ as
	\begin{align}
		\Phi^{(n)}_{s_{ij}^{(AB)}s_{kl}^{(CD)}\cdots ,\zbd}&=\left(\frac{1}{m_n\sqrt{1+z_{i}'^{(A)}}}\right)^{(1-s_{ij}^{(AB)})/2}\left(\frac{1}{m_n\sqrt{1+z_{k}'^{(C)}}}\right)^{(1-s_{kl}^{(CD)})/2}\cdots
		\notag
		\\
		&\hspace{3cm}\times
		\left[\left(Q_{i}'^{(A)}\right)^{(1-s_{ij}^{(AB)})/2}\left(Q_{k}'^{(C)}\right)^{(1-s_{kl}^{(CD)})/2}\cdots\right]
		\Phi^{(n)}_{++\cdots,\zbd}\,.
		\label{eq:multiplet}
	\end{align}

\section{Example}
\label{sec:exapmle}
In this section, we will confirm that the $\Ncal$-extended QM SUSY given in the previous section can be realized in higher dimnsional Dirac action with curved extra dimensions.
As an example, we examine the $S^2$-extra dimension with the Wu-Yang magnetic monopole background.

\subsection{Spin-weighted spherical harmonics }
As is well known, the mode functions on $S^2$-can be expressed by the spin-weighted spherical harmonics~\cite{Newman:1966ub,Torres_del_Castillo_2003,Dohi:2014fqa}.
Thus, we briefly review the Newman-Penrose $\eth$ (eth) formalism and this function.

First, we consider the rotation of the orthogonal basis $\ebd_{\theta}\,,\ebd_{\phi}$ defined in tangent space on $S^2$  
	\begin{align}
		\begin{cases}
			\ebd_{\theta}\to \ebd'_{\theta}=\ebd_{\theta}\cos\alpha-\ebd_\phi\sin\alpha\,,\\
			\ebd_{\phi}\to \ebd'_{\phi}=\ebd_{\theta}\sin\alpha+\ebd_\phi\cos\alpha\,.
		\end{cases}
	\end{align}
We call that a quantity $\eta$ has spin weight $s$, if $\eta$ transforms as follows under the above transformation:
	\begin{align}
		\eta\to e^{is\alpha}\eta\,.
	\end{align}

Furthermore, we introduce the operators $\eth$ (eth) and $\bar\eth$ (eth bar) which act as follows for $\eta$ with the spin weight $s$:
	\begin{align}
		\eth\eta&=-\left[\partial_\theta+\frac{i}{\sin\theta}\partial_\phi-s\cot\theta\right]\eta=-(\sin\theta)^s\left[\partial_\theta+\frac{i}{\sin\theta}\partial_\phi\right](\sin\theta)^{-s}\eta\,,
		\label{eq:def_eth}
		\\
		\bar\eth\eta&=-\left[\partial_\theta-\frac{i}{\sin\theta}\partial_\phi+s\cot\theta\right]\eta=-(\sin\theta)^{-s}\left[\partial_\theta-\frac{i}{\sin\theta}\partial_\phi\right](\sin\theta)^s\eta\,.
		\label{eq:def_bareth}
	\end{align}
We can show that $\eth\eta$ has spin weight $s+1$ and $\bar\eth\eta$ has spin weight $s-1$.
Therefore, $\eth$ and $\bar\eth$ correspond to the spin weight raising and lowering operators, respectively.

From \eqref{eq:def_eth} and \eqref{eq:def_bareth}, we obtain
	\begin{align}
		&\bar\eth\eth\eta=\left[
		\frac{1}{\sin\theta}\partial_\theta\sin\theta\,\partial_\theta+\frac{1}{\sin^2\theta}\partial^2_\phi+2is\frac{\cos\theta}{\sin^2\theta}\partial_\phi-\frac{s^2}{\sin^2\theta}+s(s+1)
		\right]\eta\,,
		\notag
		\\
		&\eth\bar\eth\eta=\left[
		\frac{1}{\sin\theta}\partial_\theta\sin\theta\,\partial_\theta+\frac{1}{\sin^2\theta}\partial^2_\phi+2is\frac{\cos\theta}{\sin^2\theta}\partial_\phi-\frac{s^2}{\sin^2\theta}+s(s-1)
		\right]\eta\,.
		\label{eq:ethethbar}
	\end{align}
 The spin-weighted spherical harmonics $_sY_{jm}$ with the spin weight $s$ is given as the eigenfunction of $\bar\eth\eth$ and $\eth\bar\eth$\,:
	\begin{align}
		\bar\eth\eth\,_sY_{jm}&=-\{j(j+1)-s(s+1)\}\,_sY_{jm}\,,&\eth\bar\eth\,_sY_{jm}&=-\{j(j+1)-s(s-1)\}\,_sY_{jm}\,,
		\label{eq:eigen_equation_spin-weighted_spherical_harmonics}
	\end{align}
or equivalently
	\begin{align}
		\left[
		-\frac{1}{\sin\theta}\partial_\theta\sin\theta\,\partial_\theta-\frac{1}{\sin^2\theta}\partial^2_\phi-2is\frac{\cos\theta}{\sin^2\theta}\partial_\phi+\frac{s^2}{\sin^2\theta}
		\right]\,_sY_{jm}=j(j+1)\,_sY_{jm}\,,
	\end{align}
where the spin weight is given by $s=0\,,\pm1/2\,,\pm1\,,\pm3/2\cdots$. The index $j\ (=|s|\,,|s|+1\,,|s|+2\,,\cdots)$ denotes the main total angular momentum quantum number and the index $m\ (=-j\,,-j+1\,,\cdots\,,j-1\,,j)$ indicates the secondary total angular momentum quantum number. Since the spin-weighted spherical harmonics $_sY_{jm}$ form a complete set for the fixed spin weight $s$, any function on $S^2$ with the spin weight $s$ can be decomposed into $_sY_{jm}$.

The explicit form of the normalized spin-weighted spherical harmonics is written as 
	\begin{align}
		_sY_{jm}(\theta,\phi)&=e^{im\pi}\sqrt{\frac{2j+1}{4\pi}(j+m)!(j-m)!(j+s)!(j-s)!}\notag\\
		&\hspace{2cm}\times\sum_{k=\max\{0,-m-s\}}^{\min\{j-s,j-m\}}
		\frac{(-1)^k\left(\sin\frac{\theta}{2}\right)^{m+s+2k}\left(\cos\frac{\theta}{2}\right)^{2j-m-s-2k}}
		{k!(j-m-k)!(j-s-k)!(m+s+k)!}e^{im\phi}\,,
		\label{eq:normalized_spin-weighted_spherical_harmonics}
	\end{align}
which satisfies the orthonormal relation
	\begin{align}
			\int d\Omega\,(_sY_{jm}(\theta,\phi))^*\, _sY_{j'm'}(\theta,\phi)=\delta_{jj'}\delta_{mm'}\,.
	\end{align}
Here, we have chosen the phase of this function in such a way that the function satisfies
	\begin{align}
		\eth\, _sY_{jm}=\sqrt{j(j+1)-s(s+1)}\, _{s+1}Y_{jm}\,, && \bar\eth\, _sY_{jm}=-\sqrt{j(j+1)-s(s-1)}\, _{s-1}Y_{jm}\,.
		\label{eq:eth_spherical_harmonics}
	\end{align}
In the case of $s=0$, the spin-weighted spherical harmonics corresponds to the spherical harmonic $Y_{jm}$\,.
	
As well as the ordinary spherical harmonics, this function corresponds to the representation of su(2) algebra
	\begin{align}
		L^2{_s}Y_{jm}=j(j+1)\,{_s}Y_{jm}\,,
		&&
		L_z\,{_s}Y_{jm}=m\,{_s}Y_{jm}\,,
		&&
		L_{\pm}\,{_s}Y_{jm}=\sqrt{(j\mp m)(j+1\pm m)}\,{_s}Y_{jm\pm1}\,,
		\label{eq:su(2)_rep}
	\end{align}
where $L^2,L_z$ and $L_\pm$ are the angular momentum operators for quantities with spin weight $s$
	\begin{align}
		L_z&=-i\partial_\phi\,,
		\label{eq:angular_momentum_z}
		\\
		L_\pm&=e^{\pm i\phi}\left(\pm\partial_\theta+i\cot\theta\partial_\phi-\frac{s}{\sin\theta}\right)\,,
		\\
		L^2&=\frac{1}{2}(L_+L_-+L_-L_+)+L_z^2
		=-\frac{1}{\sin\theta}\partial_\theta\sin\theta\,\partial_\theta-\frac{1}{\sin^2\theta}\partial^2_\phi-2is\frac{\cos\theta}{\sin^2\theta}\partial_\phi+\frac{s^2}{\sin^2\theta}\,,
	\end{align}
and satisfy
	\begin{align}
		[L_z\,,L_\pm]=\pm L_\pm\,,
		&&
		[L_+\,,L_-]=2L_z\,,
		&&
		[L^2\,,L_z]=[L^2\,,L_\pm]=0\,.
		\label{eq:su(2)_algebra}
	\end{align}
Furthermore, this function satisfies the following properties:
	\begin{align}
		&{_{s}}Y_{j\,m}(\pi-\theta,\phi)=(-1)^{j-m}{_{-s}}Y_{j\,m}(\theta,\phi)\,,
		\label{eq:Rphi}
		\\
		&{_s}Y_{jm}(\theta,-\phi)=(-1)^{m-s} {_{-s}}Y_{j\,-m}(\theta,\phi)\,.
		\label{eq:Rtheta}
	\end{align}

\subsection{KK mode functions and mass spectrum of $S^2$-extra dimension with magnetic monopole}
Then, we discuss the KK mode functions and the  mass spectrum of $S^2$-extra dimension with a magnetic monopole.
We consider the space $M^4\times S^2$ with the radius $a$ 
	\begin{align}
		ds^2=\eta_{\mu\nu}dx^\mu dx^\nu+a^2(d\theta^2+\sin\theta\,d\phi^2)\,,
	\end{align}
and we choose the basis of the vielbein as 
	\begin{align}
		{e_K}^{\hat K}&={\rm diag}(1,1,1,1,a,a\sin\theta)&(K=0,1,2,3,\theta,\phi\,,\ \hat K=\hat0,\hat1,\hat2,\hat3,\hat{\theta},\hat{\phi})\,.
	\end{align}
Furthermore, we introduce the Wu-Yang magnetic monopole background
	\begin{align}
		A^{N/S}
		&=-\frac{n}{2q}(\cos\theta\mp1)d\phi\ \ \ (n=0,\pm1,\pm2,\cdots)\,,
	\end{align}
where $q$ is the gauge coupling constant.\footnote{In the case of $n=0$, the Einstein equation leads to $a\to\infty$. However, in the case of $n\neq0$, the radius $a$ is stabilized and given by $a^2=n^2\kappa^2/8q^2$ where $\kappa$ is the 6D gravitational coupling constant~\cite{RandjbarDaemi:1982hi}. In this paper, we concentrate the structure of the mass spectrum of this model and we will not take into account the stability of $a$.} The gauge fields $A^N$ and $A^S$ are defined on the north patch $(0\leq\theta<\pi\,,\ 0\leq\phi<2\pi)$ and the south patch $(0<\theta\leq\pi\,,\ 0\leq\phi<2\pi)$ on $S^2$, respectively.  

Then, we consider the 6D Dirac action with the monopole background and the bulk mass $M$:
	\begin{align}
		S=\int d^4x\int_{S^2} d\theta d\phi\,a^2\sin\theta\, \bar\Psi(x,y)\left[i\Gamma^{\hat{K}}{e_{\hat{K}}}^K\left(\nabla_K+iqA_K(y)\right)-M\right]\Psi(x,y)\,,
	\end{align}
where we require that the Dirac fields on the north patch $\Psi^N(x,\theta,\phi)$ and on the south patch $\Psi^S(x,\theta,\phi)$ are related by gauge transformation
	\begin{align}
		\Psi^S=e^{-in\phi}\Psi^N\,.
	\end{align}
Here, we define the internal chiral matrix $\gamma^{in}$ which satisfies
	\begin{align}
		\gamma^{in}=i\gamma^{\hat{\theta}}\gamma^{\hat{\phi}}\,,&&
		(\gamma^{in})^2=\one_{2^{\lfloor d/2\rfloor}}\,,&&
		(\gamma^{in})^\dagger=\gamma^{in}\,,&&
		\{\gamma^{in},\gamma^{\hat{y}}\}=0\,,
	\end{align}
and we decomposition $\Psi^{N/S}(x,\theta,\phi)$ as
	\begin{align}
		\Psi^{N/S}(x,\theta,\phi)
		=\sum_{k}\sum_{\alpha=\pm}
		\Big[\,\boldsymbol{f}_{\alpha}^{(k)N/S}(\theta,\phi)\otimes\psi_{R,\alpha}^{(k)}(x)
		+\boldsymbol{g}_{\alpha}^{(k)N/S}(\theta,\phi) 
		\otimes\psi_{L,\alpha}^{(k)}(x) \,\Big]\,,
	\end{align}
where the index $\alpha=\pm$ indicate the eigenvalues of $\gamma^{in}=\pm$ for $\boldsymbol{f}_{\alpha}^{(k)N/S}$ and massless $\boldsymbol{g}_{\alpha}^{(k)N/S}$ (if exist). \footnote{The massive mode functions $\boldsymbol{g}_{\alpha}^{(k)N/S}\propto\ A\boldsymbol{f}_{\alpha}^{(k)N/S}$ are not the eigenfunction with $\gamma^{in}=\alpha$  because $A$ and $\gamma^{in}$ do not commute with each other.}
	
In this model, the operator $\Acal$ and $\Acal^\dagger$ in the supercharge \eqref{eq:supercharge} can be written into the form
	\begin{align}
		\Acal
		&=e^{\mp i\frac{n}{2}\phi}\Bigg[i\gamma^{\hat{\theta}}\frac{1}{a}\left(\eth_{s_+} P_++\bar\eth_{s_-} P_-\right)+M\Bigg]e^{\pm i\frac{n}{2}\phi}\,,
		\\
		\Acal^\dagger
		&=e^{\mp i\frac{n}{2}\phi}\Bigg[
			-i\gamma^{\hat{\theta}}\frac{1}{a}\left(\eth_{s_+} P_++\bar\eth_{s_-} P_-\right)+M
			\Bigg]e^{\pm i\frac{n}{2}\phi}\,,
	\end{align}
where the upper and lower signs in the exponential denote the ones for the north and south patches, respectively. (Here after, we use the same notation for the signs in the exponential if appear.)
The operators $\eth_{s_+}$ and $\bar\eth_{s_-}$ represent the spin weight raising and lowering operators for the spin weight $s_\pm=-(n\pm1)/2$, and $P_\pm=(1\pm\gamma^{in})/2$ are the projection matrices for $\gamma^{in}$.
Then, $\Acal^\dagger \Acal$ and $\Acal\Acal^\dagger$ are given by
	\begin{align}
		\Acal^\dagger \Acal&=\Acal\Acal^\dagger
		=e^{\mp i\frac{n}{2}\phi}\Bigg[
			-\frac{1}{a^2}\left(\bar\eth_{s_++1}\eth_{s_+} P_++\eth_{s_--1}\bar\eth_{s_-} P_-\right)+M^2
			\Bigg]e^{\pm i\frac{n}{2}\phi}\,.
		\label{eq:AAdagger}
	\end{align}
Thus, we can find that the mode functions and the mass eigenvalues are obtained as follows from \eqref{eq:SUSY_relation} and \eqref{eq:eigen_equation_spin-weighted_spherical_harmonics}:
	\begin{align}
		\boldsymbol{f}_{\alpha}^{(jm)N/S}(\theta,\phi)=\frac{{_{s_\alpha}}Y_{jm}(\theta,\phi)\,e^{\mp i\frac{n}{2}\phi}}{a}\ebd_{\alpha}\,,
		&&
		\boldsymbol{g}_{\alpha}^{(jm)N/S}(\theta,\phi)=\frac{1}{m_{j}}\Acal\boldsymbol{f}_{\alpha}^{(jm)N/S}(\theta,\phi)\,,
		&&
		(\alpha=\pm)\,,
		\\
		m_{j}=\sqrt{\frac{j(j+1)}{a^2}-\frac{n^2-1}{4a^2}+M^2}\,,
	\end{align}
where $s_\pm$, $j$ and $m$ are given by
	\begin{align}
		s_\pm=-\frac{n\pm1}{2}\,,
		&&
		j=
		\begin{cases}
			|s_+|\,,|s_+|+1\,,|s_+|+2\,,\cdots & (\text{for}\ \alpha=+)
			\\
			|s_-|\,,|s_-|+1\,,|s_-|+2\,,\cdots & (\text{for}\ \alpha=-)
		\end{cases}\,,
		&&
		m=-j\,,-j+1\,,\cdots\,,j-1\,,j\,,
	\end{align}
and we define $\ebd_{\pm}$ as the  2-component orthonormal vectors which satisfy 
	\begin{align}
		\gamma^{in}\ebd_{\pm}=\pm\ebd_{\pm}\,, 
		&& 
		\gamma^{\hat{\theta}}\ebd_{\pm}=i\ebd_{\mp}\,,
		&&
		\gamma^{\hat{\phi}}\ebd_{\pm}=\mp\ebd_{\mp}\,.
	\end{align}
Then, the degeneracy of $j$-th KK level is $2j+1$ for each mode function.

We can see that the mode functions $\boldsymbol{f}_{\pm}^{(jm)N/S}$ and $\boldsymbol{g}_{\pm}^{(jm)N/S}$ with $j=|s_\pm|$ have the lowest mass eigenvalue $\sqrt{1/a^2+M^2}$ in the case  without the monopole $(n=0)$.
On the other hand, in the case with the monopole $(n\neq0)$, the lowest mass eigenvalue decrease to $M$ and the corresponding mode functions are given by $\boldsymbol{f}_{-}^{(|s_-|\,m)N/S}$ and $\boldsymbol{g}_{-}^{(|s_-|\,m)N/S}$ for $n>0$ and $\boldsymbol{f}_{+}^{(s_+\,m)N/S}$  and $\boldsymbol{g}_{+}^{(s_+\,m)N/S}$ for $n<0$.
Thus, there are no massless modes unless the bulk mass $M$ equals to 0 and the magnetic monopole exists.
If we consider the model with $M=0$ and $n>0\ (n<0)$, $|n|$ massless modes exist for the mode functions $\boldsymbol{f}_{\alpha}^{(jm)N/S}$ and $\boldsymbol{g}_{\alpha}^{(jm)N/S}$ with $\alpha=-\ (\alpha=+)$ respectively, and $|n|$ massless 4D Dirac fields appear.\footnote{If we consider the action with a 6D chiral field with the eigenvalue $\Gamma^7=-\Gamma^{\hat{0}}\Gamma^{\hat{1}}\Gamma^{\hat{2}}\Gamma^{\hat{3}}\Gamma^{\hat{\theta}}\Gamma^{\hat{\phi}}=+\ (-)$, there appear $n$ 4D left (right)-handed fields in the case of $n>0$ and $|n|$ 4D right (left)-handed fields in the case of $n<0$.
}

\subsection{$\Ncal$-extended QM SUSY in $S^2$-extra dimension with magnetic monopole}

Let us construct the $\Ncal$-extended QM SUSY in $S^2$-extra dimension with the magnetic monopole. For this purpose, we first summarize the well defined operators for the KK mode functions $\boldsymbol{f}_{\alpha}^{(jm)N/S}$ and their properties in this model:
\\\\
{\bf $\bullet$ properties of} $\gamma^{in}\gamma^{\hat{y}}\eth_{in}$

We write the operator $\eth_{in}$ as 
	\begin{align}
		\eth_{in}=e^{\mp i\frac{n}{2}\phi}
		(\eth_{s_+} P_++\bar\eth_{s_-} P_-)e^{\pm i\frac{n}{2}\phi}\,,
	\end{align}
 which appears in the $\Acal$ and $\Acal^\dagger$.
Then, the operators $i\gamma^{in}\gamma^{\hat{y}}\eth_{in}\ (\hat{y}=\hat{\theta}\,,\hat{\phi})$ satisfy
	\begin{align}
		[i\gamma^{in}\gamma^{\hat{y}}\eth_{in}\,,\Acal^\dagger\Acal]=0\,,
		&&
		\{i\gamma^{in}\gamma^{\hat{\theta}}\eth_{in}\,,i\gamma^{in}\gamma^{\hat{\phi}}\eth_{in}\}=0\,.
	\end{align}
These operators relate  $\boldsymbol{f}_{\alpha}^{(jm)N/S}$ with $\boldsymbol{f}_{-\alpha}^{(jm)N/S}$ since the operator $\eth_{in}$ raises the spin weight for the functions with $\gamma^{in}=+$ and lowers the spin weight for the functions with $\gamma^{in}=-$. Furthermore, these are Hermitian for $\boldsymbol{f}_{\alpha}^{(jm)N/S}$ in the sence that
	\begin{align}
		\langle i\gamma^{in}\gamma^{\hat{y}}\eth_{in}\boldsymbol{f}_{\alpha'}^{(j'm')N/S}|\boldsymbol{f}_{\alpha}^{(jm)N/S}\rangle=\langle \boldsymbol{f}_{\alpha'}^{(j'm')N/S}|i\gamma^{in}\gamma^{\hat{y}}\eth_{in}\boldsymbol{f}_{\alpha}^{(jm)N/S}\rangle\,.
	\end{align}

We can use these operators to construct the $\Ncal$-extended QM SUSY.
\\\\
{\bf$\bullet$  properties of angular momentum operator}

From the rotational symmetries of $S^2$-extra dimension, the angular momentum operator $L_z$ given in \eqref{eq:angular_momentum_z} with gauge transformation
	\begin{align}
		L^{(n)}_z=e^{\mp i\frac{n}{2}\phi}L_ze^{\pm i\frac{n}{2}\phi}\,, 
	\end{align}
are Hermitian and well defined for the mode functions. Furthermore this satisfy
	\begin{align}
		[L^{(n)}_z\,, \Acal^\dagger\Acal]=[L^{(n)}_z\,, \eth_{in}]=0\,.
	\end{align}
This is also useful for the construction of the $\Ncal$-extended QM SUSY.
\\\\
{\bf$\bullet$  properties of reflection operators}

In the model without the magnetic monopole ($n=0$), we can consider the reflection operators with gamma matrices $\gamma^{in}\gamma^{\hat{\theta}}R_{\theta}$ and $\gamma^{in}\gamma^{\hat{\phi}}R_{\phi}$ 
where $R_{\theta}$ and $R_{\phi}$ represent the reflections for the coordinates $\theta$ and $\phi$
\begin{align}
R_{\theta}: \theta\to\pi-\theta\,,
&&
R_{\phi}: \phi\to-\phi\,.
\end{align}
These operators are well defined for the mode functions $\boldsymbol{f}_{\alpha}^{(jm)}$ and commute with $\Acal^\dagger\Acal$
because of the reflection symmetries in $S^2$-extra dimension. 

In the case of $S^2$ with monopole ($n\neq 0$), the above operators are ill defined and do not commute with $\Acal^\dagger\Acal$ by the monopole background.
However, the operator with the combination of the above reflections and the gauge transformation
	\begin{align}
		\Rcal=e^{\mp i\frac{n}{2}\phi}R_\theta R_\phi e^{\pm i\frac{n}{2}\phi}
	\end{align}
is Hermitian and well defined for the mode functions  and satisfies
	\begin{align}
		&[\Rcal\,,\Acal^\dagger\Acal]=\{\Rcal\,,\eth_{in}\}=\{\Rcal\,,L^{(n)}_z\}=0\,,
		&\Rcal^2=\one\,.
	\end{align}
We find that this operator connects the mode functions $\boldsymbol{f}_{\alpha}^{(jm)N/S}$ to $\boldsymbol{f}_{\alpha}^{(j-m)N/S}$.

Then, we can obtain various kinds of $\Ncal$-extended QM SUSYs from the above operators.
As examples, we consider the following two $\Ncal$-extended QM SUSYs:
\\\\
{\bf$\bullet$  $\Ncal=6$ extended QM SUSY with central charges from the angular momentum operator}

First, we consider the $\Ncal=6$ extended QM SUSY with central charges which consist of the angular momentum operator $L_z^{(n)}$:
	\begin{align}
		\{Q_k\,,Q_l\}&=2H\delta_{kl}+2Z_{kl}\,,
		\label{eq:ex_susy_alg}
		\\
		[Q_{k'}\,,Z_{kl}]&=[H\,,Z_{kl}]=[Z_{kl}\,,Z_{k'l'}]=[Q_{k'}\,,H]=0\ \ (k,l,k',l'=1,\cdots ,6)\,,
	\end{align}
where the supercharges $Q_k$ and the nonzero components of central charges $Z_{kl}$ are given by
	\begin{align}
		Q_1&=\begin{pmatrix}
		0 &  -\gamma^{in}\gamma^{\hat{\theta}}\eth_{in}\Acal^\dagger\\
		\Acal \gamma^{in}\gamma^{\hat{\theta}}\eth_{in} & 0
		\end{pmatrix}
		\,,
		&
		Q_2&=\begin{pmatrix}
		0 &  -\gamma^{in}\gamma^{\hat{\phi}}\eth_{in}\Acal^\dagger\\
		\Acal \gamma^{in}\gamma^{\hat{\phi}}\eth_{in} & 0
		\end{pmatrix}
		\,,
		\\
		Q_3&=\begin{pmatrix}
		0 & i\gamma^{in}L^{(n)}_z\Acal^\dagger \\ -i\Acal\gamma^{in}L^{(n)}_z & 0
		\end{pmatrix}\,,
		&
		Q_4&=\begin{pmatrix}
		0 & i\gamma^{in}\Acal^\dagger \\ -i\Acal\gamma^{in} & 0
		\end{pmatrix}
		\,,
		\\
		Q_5&=\begin{pmatrix}
		0 & L^{(n)}_z\Acal^\dagger \\ \Acal L^{(n)}_z & 0
		\end{pmatrix}\,,
		&
		Q_6&=Q\,,
		\\
		Z_{11}&=Z_{22}
		=H(-1+a^2H-a^2M^2)\,,
		\\
		Z_{34}&=Z_{56}=\begin{pmatrix}
		L^{(n)}_z\Acal^\dagger\Acal & 0 \\ 0 & \Acal L^{(n)}_z\Acal^\dagger
		\end{pmatrix}\,,
		&
		Z_{33}=Z_{55}&=-H+
		\begin{pmatrix}
		(L^{(n)}_z)^2\Acal^\dagger\Acal & 0 \\ 0 & \Acal (L^{(n)}_z)^2\Acal^\dagger
		\end{pmatrix}\,,
	\end{align}
In this QM SUSY, $Q_3$ and $Q_4$ commute with each other and also $Q_5$ and $Q_6$.

According to the discussion given in the section \ref{subsec:rep_ex_SUSY}, we redefine the supercharges for the eigenfunctions with the eigenvalues $H=m_j^2$,\ $Z_{34}=m\,m_j^2$ and $Z_{33}=(-1+m^2)\,m_j^2$ (where $m$ indicates the angular momentum number):
	\begin{align}
		Q_1'&=Q_1\,,
		&
		Q_2'&=Q_2\,,
		\\
		Q_3'&=\frac{1}{\sqrt{1+m^2}}\left(mQ_3+Q_4\right)\,,
		&
		Q_4'&=\frac{1}{\sqrt{1+m^2}}\left(Q_3-mQ_4\right)\,,
		\\
		Q_5'&=\frac{1}{\sqrt{1+m^2}}\left(mQ_5+Q_6\right)\,,
		&
		Q_6'&=\frac{1}{\sqrt{1+m^2}}\left(Q_5-mQ_6\right)\,,
	\end{align}
The SUSY algebra \eqref{eq:ex_susy_alg}  is written into the following form for the redefined supercharges
	\begin{align}
			\{Q_k'\,,Q_l'\}&=2m_j^2\delta_{kl}+2z'_{k}m_j^2\delta_{kl}\,,
	\end{align}
where $z'_{k}\ (k=1,\cdots,6)$ is given by
	\begin{align}
		z'_{1}=z'_{2}=(-1+a^2m_j^2-a^2M^2)\,,
		&&
		z'_{3}=z'_{5}=m^2\,,
		&&
		z'_{4}=z'_{6}=-1\,.
		\label{eq:diagonalized_CC}
	\end{align}
From this relation, we find that $Q_4'=Q_6'=0$ for any eigenstates. Furthermore the supercharges $Q_1'$ and $Q_2'$ equal to 0 for the states with $m_j=M$ (or equivalently $H=M^2$), that is the case of $j=|s_-|$ for $n>0$ and $j=s_+$ for $n<0$. Thus, the BPS states appear as discussed in the section \eqref{sec:N-extended}.

Then we can construct the eigenfunctions $\Phi^{(jm)}_{s_{12}s_{35},\zbd'}(y)$ with {\color{red}$H=m_j^2\neq M^2$} as 
	\begin{align}
		\Phi^{(jm)N/S}_{++,\zbd'}=	\begin{pmatrix}
		\boldsymbol{f}_{+}^{(jm)N/S} \\ 0
		\end{pmatrix}\,,
		&&
		\Phi^{(jm)N/S}_{-+,\zbd'}=\begin{pmatrix}
		0 \\ -i\boldsymbol{g}_{-}^{(jm)N/S} 
		\end{pmatrix}\,,
		&&
		\Phi^{(jm)N/S}_{+-,\zbd'}=\begin{pmatrix}
		0 \\ -i\boldsymbol{g}_{+}^{(jm)N/S} 
		\end{pmatrix}\,,
		&&
		\Phi^{(jm)N/S}_{--,\zbd'}=\begin{pmatrix}
		\boldsymbol{f}_{-}^{(jm)N/S} \\ 0
		\end{pmatrix}\,,
	\end{align}
and also the eigenfunctions $\Phi^{(jm)}_{s_{35},\zbd'}(y)$ with {\color{red}$H=M^2$} as
	\begin{align}
		\begin{cases}
		\Phi^{(|s_-|\,m)N/S}_{+,\zbd'}=	\begin{pmatrix}
		0 \\ -i\boldsymbol{g}_{-}^{(|s_-|\,m)N/S}
		\end{pmatrix}\,,
		\ \ \ 
		\Phi^{(|s_-|\,m)N/S}_{-,\zbd'}=\begin{pmatrix}
		-\boldsymbol{f}_{-}^{(|s_-|\,m)N/S} \\ 0 
		\end{pmatrix}
		&\text{for}\ n>0\ \text{and}\ j=|s_-|\,,
		\\
		\Phi^{(s_+m)N/S}_{+,\zbd'}=	\begin{pmatrix}
		\boldsymbol{f}_{+}^{(s_+m)N/S} \\ 0
		\end{pmatrix}\,,
		\ \ \ 
		\Phi^{(s_+m)N/S}_{-,\zbd'}=\begin{pmatrix}
		0 \\ -i\boldsymbol{g}_{+}^{(s_+m)N/S} 
		\end{pmatrix}
		&\text{for}\ n<0\ \text{and}\ j=s_+\,.
		\end{cases}
	\end{align}	
Here $\zbd'$ indicates $\{z_i'\ (i=1,\cdots,6)\}$ given in \eqref{eq:diagonalized_CC} and $s_{12}=\pm$, $s_{35}=\pm$ denote the signs of the eigenvalues of $S_{12}=-iQ_1Q_2$, $S_{35}=-iQ_3Q_5$.

Then we can show that the above eigenfunctions form the supermultiplets and satisfy the same relation as \eqref{eq:multiplet}.
In this extended QM SUSY, the four (two) -fold degenerated eigenfunctions with {\color{red}$H\neq M^2\ (H=M^2)$} are related by the supercharges $Q_k'$ with $k=1,2,3,5$\ ($k=3,5$).
Furthermore, the extra $(2j+1)$-fold degeneracy exists in the eigenfunctions.
 The origin of this degeneracy comes from the angular momentum number $m$, which corresponds to the eigenvalues of central charges.
\\\\\\
{\bf$\bullet$  $\Ncal=6$ extended QM SUSY with central charges from the angular momentum operator and the reflection operator}

Next, we construct the $\Ncal=6$ extended QM SUSY with central charges from the angular momentum operator and the reflection operator:
	\begin{align}
		\{Q_k\,,Q_l\}&=2H\delta_{kl}+2Z_{kl}\,,
		\notag
		\\
		[Q_{k'}\,,Z_{kl}]&=[H\,,Z_{kl}]=[Z_{kl}\,,Z_{k'l'}]=[Q_{k'}\,,H]=0\ \ (k,l,k',l'=1,\cdots ,6)\,,
	\end{align}
where the supercharges $Q_k$ and the nonzero components of the central charges are taken to be the form of
	\begin{align}
		Q_1&=\begin{pmatrix}
		0 &  -\gamma^{in}\gamma^{\hat{\theta}}\eth_{in}\Acal^\dagger\\
		\Acal \gamma^{in}\gamma^{\hat{\theta}}\eth_{in} & 0
		\end{pmatrix}
		\,,
		&
		Q_2&=\begin{pmatrix}
		0 &  -\gamma^{in}\gamma^{\hat{\phi}}\eth_{in}\Acal^\dagger\\
		\Acal \gamma^{in}\gamma^{\hat{\phi}}\eth_{in} & 0
		\end{pmatrix}
		\,,
		\\
		Q_3&=\begin{pmatrix}
		0 & i\gamma^{in}L^{(n)}_z\Acal^\dagger \\ -i\Acal\gamma^{in}L^{(n)}_z & 0
		\end{pmatrix}
		\,,
		&
		Q_4&=\begin{pmatrix}
		0 & -L^{(n)}_z\Rcal\Acal^\dagger \\ \Acal L^{(n)}_z\Rcal & 0
		\end{pmatrix}
		\,,
		\\
		Q_5&=\begin{pmatrix}
		0 & i\Rcal\Acal^\dagger \\ -i\Acal\Rcal & 0
		\end{pmatrix}
		\,,
		&
		Q_6&=Q\,,
		\\
		Z_{11}&=Z_{22}
		=H(-1+a^2H-a^2M^2)\,,
		&
		Z_{33}&=Z_{44}=-H+
		\begin{pmatrix}
		(L^{(n)}_z)^2\Acal^\dagger\Acal & 0 \\ 0 & \Acal (L^{(n)}_z)^2\Acal^\dagger
		\end{pmatrix}\,,
	\end{align}
As well as the previous extended QM SUSY, the supercharges $Q_1$ and $Q_2$ equal to 0 for the states with $H=M^2$. In addtion, the eigenstates with $Z_{33}=-m_j^2\ (L_z^{(n)}=0)$ become BPS states and $Q_3$ and $Q_4$ equal to 0 for them.  

Here, we redefine the basis of the KK mode functions to obtain the representation of this QM SUSY
	\begin{align}
		\boldsymbol{f}_{\alpha,r}^{(jm)N/S}(\theta,\phi)&=\frac{1}{\sqrt2}\left(\boldsymbol{f}_{\alpha}^{(jm)N/S}(\theta,\phi)+r(-1)^{j+s_\alpha}\boldsymbol{f}_{\alpha}^{(j-m)N/S}(\theta,\phi)\right)\,,
		\notag
		\\
		\boldsymbol{g}_{\alpha,r}^{(jm)N/S}(\theta,\phi)&=\frac{1}{\sqrt2}\left(\boldsymbol{g}_{\alpha}^{(jm)N/S}(\theta,\phi)+r(-1)^{j+s_\alpha}\boldsymbol{g}_{\alpha}^{(j-m)N/S}(\theta,\phi)\right)\,,
		\label{eq:redef_mode_function}
	\end{align}
where $r=\pm$ and $m\geq 0$. The mode functions $\boldsymbol{f}_{\alpha,r}^{(jm)N/S}$ with $r=\pm$ correspond to the parity even and odd functions for $\Rcal$, respectively
	\begin{align}
		\Rcal\boldsymbol{f}_{\alpha,\pm}^{(jm)N/S}=\pm\boldsymbol{f}_{\alpha,\pm}^{(jm)N/S}\,.
	\end{align}
Then, we can obtain the eigenfunctions $\Phi^{(jm)}_{s_{12}s_{34}s_{56},\zbd}(y)$ for non-BPS states ($H=m_j^2\neq M^2$, $Z_{33}=(-1+m^2)m_j^2\neq -m_j^2$) as follows:
	\begin{align}
		\Phi^{(jm)}_{+++,\zbd}=	\begin{pmatrix}
		\boldsymbol{f}_{+,+}^{(jm)N/S} \\ 0
		\end{pmatrix}\,,
		&&
		\Phi^{(jm)}_{-++,\zbd}=\begin{pmatrix}
		0 \\ -i\boldsymbol{g}_{-,-}^{(jm)N/S} 
		\end{pmatrix}\,,
		&&
		\Phi^{(jm)}_{+-+,\zbd}=\begin{pmatrix}
		0 \\ -i\boldsymbol{g}_{+,-}^{(jm)N/S} 
		\end{pmatrix}\,,
		&&
		\Phi^{(jm)}_{++-,\zbd}=\begin{pmatrix}
		0\\ -i\boldsymbol{g}_{+,+}^{(jm)N/S} 
		\end{pmatrix}\,,
		\notag
		\\
		\Phi^{(jm)}_{--+,\zbd}=	\begin{pmatrix}
		\boldsymbol{f}_{-,+}^{(jm)N/S} \\ 0
		\end{pmatrix}\,,
		&&
		\Phi^{(jm)}_{-+-,\zbd}=\begin{pmatrix}
		\boldsymbol{f}_{-,-}^{(jm)N/S} \\ 0
		\end{pmatrix}\,,
		&&
		\Phi^{(jm)}_{+--,\zbd}=\begin{pmatrix}
		\boldsymbol{f}_{+,-}^{(jm)N/S} \\ 0
		\end{pmatrix}\,,
		&&
		\Phi^{(jm)}_{---,\zbd}=\begin{pmatrix}
		0 \\ -i\boldsymbol{g}_{-,+}^{(jm)N/S}
		\end{pmatrix}\,,
		\label{eq:N=6_multiplet}
	\end{align}
where $\zbd$ indicates $z_k\  (k=1,2,3,5)$ which are the coefficient of $m_j^2$ in the eigenvalues of $Z_{kk}$
	\begin{align}
		z_1=z_2=-1+a^2m_j^2-a^2M^2\,,&&z_3=z_5=-1+m^2\,,
	\end{align}
and the dependence of $\zbd$ in the right hand side of \eqref{eq:N=6_multiplet} is described by $j$ and $m$.
$s_{12}=\pm$, $s_{34}=\pm$ and $s_{56}=\pm$ denote the signs of the eigenvalues of $S_{12}=-iQ_1Q_2$, $S_{34}=-iQ_3Q_4$, $S_{56}=-iQ_5Q_6$. 

We can show that the above eigenfunctions form the supermultiplets and satisfy the same relation as \eqref{eq:multiplet}.
The supermultiplets for the BPS states can be constructed by use of the mode functions $\boldsymbol{f}_{\alpha,r}^{(jm)N/S}$ and $\boldsymbol{g}_{\alpha,r}^{(jm)N/S}$ in the same way.

In this extended QM SUSY, the eight-fold degenerated non-BPS states is related by the six supercharges $Q_k\ (k=1,\cdots,6)$.
Therefore, we see that the additional two-fold degeneracy can be further explained by the supercharges compared with the previous extended QM SUSY (although the total degeneracy including the BPS states with $m=0$ is not changed).
This additional degeneracy corresponds to the parity even and odd for the reflection. Since the eigenfunctions with $m=0$ are only equivalent for the  parity even from the definition \eqref{eq:redef_mode_function}, they should become the BPS states.

Although we considered two examples, there are more 
extended QM SUSYs constructed from the symmetries.
For instance, we can obtain them by using the reflection operators $\gamma^{in}\gamma^{\hat{\theta}}R_{\theta}$ and $\gamma^{in}\gamma^{\hat{\phi}}R_{\phi}$ in the case without the magnetic monopole, since those operators correspond to the symmetries in such model.
\section{Summary and discussion}
\label{sec:summary}

In this paper, we have constructed the new realization of the $\Ncal$-extended QM SUSY with central charges hidden in the higher dimensional Dirac action with curved extra dimensions. 
This extended QM SUSY results from symmetries in extra dimensions, and the supercharges and the central charges are obtained by use of them. We have also investigated the representation of the SUSY algebra and shown that the supermultiplets would become the BPS states. 
In addition, we considered the model of $S^2$-extra dimension with the magnetic monopole as the concrete example.
Then we have confirmed that the KK mode functions 
properly correspond to the representations in the two types of $\Ncal$-extended QM SUSYs which are obtained from the rotational and reflection symmetries in the extra dimensions.

The characteristic property of our extended QM SUSY is that certain supercharges commute with each other. Recently, a new generalization of supersymmetry is proposed, that is $\mathbb{Z}_2^n$-graded supersymmetry whose supercharges have $n$ degrees~\cite{Bruce:2018tbl,Bruce:2019pjy,Aizawa:2019nza}. For their supercharges with different degrees, the algebra may not close in anticommutator but commutator. Therefore, it is interesting to investigate the relation between this supersymmetry and our QM SUSY.

Our analysis is not perfect. 
As we have seen in the section \ref{sec:exapmle}, there appear various extended QM SUSYs in a model, according to sets of symmetries.
Thus, we should clarify what kinds of extended QM SUSYs can be obtained from a model.
It is also important to reveal that what symmetries exist by the choice of boundary conditions, extra dimensional spaces and background fields since they would affect the structures of extended QM SUSYs.
Furthermore, it is known that central charges are closely related to topological properties~\cite{Witten:1978mh,Seiberg:1994aj,Seiberg:1994rs}. Although the central charges given in the section \ref{sec:exapmle} might be related to the topology of $S^2$ and the magnetic monopole, the detailed structures are not unveiled.
Thus, we should more investigate models with nontrivial topology.
We can also expect the possibilities that the further structures are hidden in the 4D mass spectrum.
Since our discussions are not completely general, other realizations of extended QM SUSY might be constructed.

Moreover, since we have obtained the new extended QM SUSY, it is fascinating to study new types of exactly solvable quantum-mechanical models. 
The issues mentioned in this section will be reported in a future work.

\section*{Acknowledgement}
The author thanks Y. Fujimoto, K. Hasegawa, K. Nishiwaki, M. Sakamoto, K. Tatsumi for discussions in the early
stage of this work.

\bibliographystyle{elsarticle-num}
\bibliography{Reference}

\end{document}